# Extension Language Automation of Embedded System Debugging

Dale Parson, Bryan Schlieder and Paul Beatty

# Demo: Exploiting Extension Language Capabilities in Hardware / Software Co-design and Co-verification

Cathy Moeller, Dale Parson, Bryan Schlieder, Jay Wilshire

*Bell Laboratories / Lucent Technologies*

# Extension Language Automation of Embedded System Debugging

Dale Parson
dparson@lucent.com

Bryan Schlieder
bryanschlieder@lucent.com

Paul Beatty
pebeatty@lucent.com

All authors:   Bell Laboratories, Lucent Technologies
1247 South Cedar Crest Blvd.
Allentown, Pa. 18103

**Abstract**

*Embedded systems contain several layers of target processing abstraction. These layers include electronic circuit, binary machine code, mnemonic assembly code, and high-level procedural and object-oriented abstractions. Physical and temporal constraints and artifacts within physically embedded systems make it impossible for software engineers to operate at a single layer of processor abstraction. The Luxdbg embedded system debugger exposes these layers to debugger users, and it adds an additional layer, the extension language layer, that allows users to extend both the debugger and its target processor capabilities. Tcl is Luxdbg's extension language. Luxdbg users can apply Tcl to automate interactive debugging steps, to redirect and to interconnect target processor input-output facilities, to schedule multiple processor execution, to log and to react to target processing exceptions, and to automate target system testing. Inclusion of an extension language like Tcl in a debugger promises additional advantages for distributed debugging, where debuggers can pass extension language expressions across computer networks.*

## 1. Introduction

Embedded system debugging involves more conceptual layers of a target system than debugging for time-sharing systems. Consider the case of debugging a C program within a time-sharing system. User-debugger interaction occurs almost entirely at a C language level of abstraction. Descent into assembly language and machine code representations of a target program is rare. Suspicions about a compiler bug may require inspection of generated assembly code. Inadvertent stepping into an optimized library subroutine leads to display of assembly mnemonics and binary numbers. Debugging concurrency problems in multi-threaded programs entails cognizance of time, but well-structured multi-threaded programs limit temporal awareness to a few, isolated regions where multiple threads are allowed to interact. These examples are exceptions, and most programmers can debug their programs exclusively from a source language perspective.

Embedded systems add several dimensions to debugging. Embedded systems include programmable physical devices that have no direct language counterparts at higher levels of abstraction. Their programming requires direct manipulation of registers and state machines. Assembly language programming is common for performance-critical modules. Temporal determinacy is fundamental to a real-time embedded system, eliminating the possibility of constraining temporal awareness to a few, isolated regions of code. Multiple, heterogeneous processors operating at different levels of abstraction, for example a microcontroller running Java[TM] or C++ teamed with one or more digital signal processors (DSPs) running a mix of C and assembly code, are commonplace within some classes of embedded systems. Mixing abstractions within design and debugging is typical.

This paper is about the practical application of an additional layer of language abstraction, an *extension language abstraction*, to the aforementioned layers of embedded systems. An extension language serves a tool such as a debugger by providing a programming language, typically an interpreted language, for extending tool capabilities via composition at tool usage time. Basic tool capabilities ultimately constrain the power of extensions that users can compose.

Luxdbg, the LUxWORKS tool suite's debugger for embedded systems [1,2], exposes multiple layers of target system abstraction to debugger users and auxiliary tools. Lucent Microelectronics provides a production version of Luxdbg in support of Lucent's digital signal processors and mixed microcontroller-DSP systems for embedded telecommunications and related applications [3]. Luxdbg is implemented in C++, and it uses Tcl as its built-in extension language [4].

Luxdbg application space has primarily been in the area of embedded telecommunications signal processing, a huge area that is seeing rapid growth in multiprocessing. Cellular basestations — the electronics connected to the towers one sees while traveling along the road — are coming to employ large banks of two- and three-core DSP chips that put hundreds of individual processing cores into a system. Customers designing circuit boards for these systems require a debugger that can connect to about one hundred processors at a time. Not all processors are being debugged at any given time, and many of the signal processing algorithms running on each processor involve only that processor. The DSPs process signals for parallel voice and data channels. Nevertheless, all processors are running at the same time, and any processor can exhibit a bug at any time, so the debugger must be capable of connecting and interacting with one or more of the one hundred at any time.

At the other end of the cellular continuum, a cellular handset (cell phone) often contains a DSP and a microcontroller. The DSP handles signal processing for the voice and data channels within that handset, while the microcontroller controls the system and provides user level IO capabilities. The handset presents only two processors, but they are heterogeneous, i.e., they provide different programming architectures and instruction sets. They may be programmed in different languages, e.g., assembly language or C for the DSP and C, C++ or Java™ for the microcontroller. Handset debugging does not require the massive multiprocessor debugging features required by a basestation, but it does require the ability to debug heterogeneous instruction sets and languages.

A detailed account of the signal processing and related control architectures of Luxdbg's target embedded systems is outside the scope of this paper, and indeed it is unnecessary. Most embedded systems consist of an admixture of high level language programming processes, assembly code for performance-critical tasks, and hardware registers and special-purpose devices for accepting sensory input and producing sensory or sensorimotor output. Communications channels can be considered artificial, special-purpose sensory realms. Most of the programming examples in this paper come from multiprocessor signal processing systems, where a given processor reads an information-bearing signal frame, processes that frame (e.g., extracts information, encodes information, superimposes signal content on a carrier, encrypts, decrypts, removes noise or otherwise filters, etc.), and then sends the resulting frame of data on to the next stage. Human users usually attach at the endpoints of these distributed signal flows. Debugger users attach anywhere a bug surfaces.

This paper constitutes an experience report in effective uses of an extension language within a multiple abstraction embedded system debugger. Section 2 gives an overview of using an extension language within an application. Section 3 examines the layers of embedded system abstraction for which Luxdbg supports debugging. Section 4 surveys the classes of extensions that users can employ in extending Luxdbg and its target embedded processors. Section 5 discusses related work. Section 6 gives conclusions and directions for upcoming work. Section 7 gives acknowledgments. Section 8 is an appendix containing a description of a Luxdbg demo that utilizes Tcl to model inter-processor communication hardware.

## 2. Extension language systems

Figure 1 diagrams the interactions of an extension language in a system. The main system components are a *client* such as a user, extension script or auxiliary tool (e.g., GUI), an *extension language interpreter* such as Tcl[4] or Python [5], and an underlying *application* such as a debugger. Interaction begins at initialization time, when the application registers a number of primitive functions with the extension language interpreter. The interpreter adds these functions to its native command set.

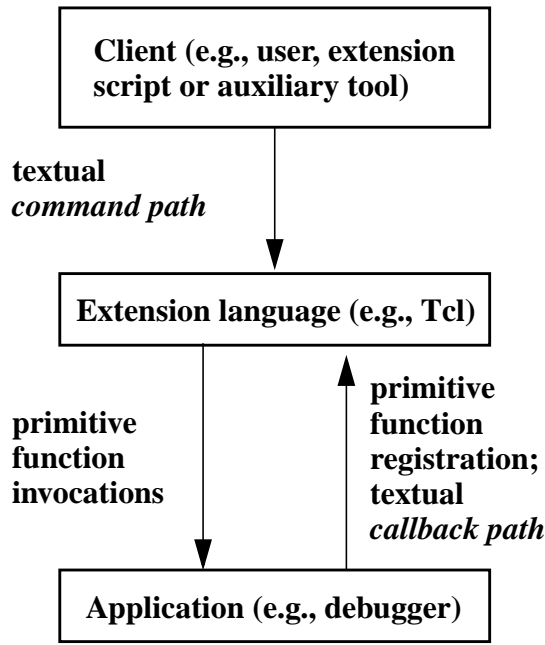

**Figure 1: Extension language interactions**

After initialization the system enters its main interaction loop. It uses a *command path* to pass client commands to the extension language interpreter. The user may enter commands via a textual or graphical interface, or the user may write and invoke an extension language program (a.k.a. *script*). Built-in extension language primitives include control flow and data structuring constructs. The extension interpreter parses commands and executes scripts by invoking both its own primitives and application primitives. Typical debugger primitives include data retrieval, data modification and execution control for target processing systems.

Figure 1 also shows a *callback path* from the application to the extension language interpreter. In an event-driven system it is possible for a user to associate an extension language expression with an event in the application layer. For example, in Luxdbg a user can associate a Tcl expression with a target processor breakpoint. When the breakpoint occurs, the debugger calls back to the Tcl interpreter, passing the expression to be evaluated and an identifier for the breakpoint as parameters. The interpreter evaluates the expression and returns the result to the debugger. Expression evaluation may include retrieval and modification of processor state. During a callback the extension language interpreter acts as a servant for the application layer, a reversal of their normal roles. Whereas the command path of Figure 1 allows the extension language to build extensions out of primitives, the callback path allows the extension language to extend primitives. For Luxdbg this means that users can extend the built-in debugging layers discussed in the next section.

## 3. Luxdbg layers of processor abstraction

Figure 2 illustrates the layers of abstraction available to a Luxdbg user. The *extension language processor layer* of Figure 2 is the extension language interpreter of Figure 1, augmented with Luxdbg debugging primitives. The remaining layers of Figure 2 constitute the application module of Figure 1. Each layer provides a C++ API that allows outer layers to build upon it. Luxdbg supports concurrent debugging of multiple, heterogeneous *virtual processors*, where each virtual processor is a hardware processor, a processor simulation model, or an operating system process that implements the layers of processing abstraction of Figure 2. Luxdbg represents each target processor as a C++ object with these layers. Each object includes public methods for access, modification, and execution control at these layers.

The *circuit layer* represents integrated circuit pins, processor registers, memory regions, peripheral devices and timing information. This is the most physical layer, closest to the hardware. Embedded system programming entails access to devices such as digital-to-analog converters that control physical devices (e.g., speakers, heaters or motors), analog-to-digital converters that allow the system to monitor sensors (e.g., microphones or temperature sensors), coder-decoders (*codecs*) that translate signals between computational and communications-oriented representations, and other devices that manipulate physical signals in a variety of ways. An embedded processor interacts with physical devices via registers that it retrieves and modifies using dedicated processor register slots and dedicated IO instructions, or by using memory-mapped registers, conventional fetch-store instructions and direct memory access (DMA). Embedded devices house state for a program that is outside the bounds of memory-oriented program variable state. An embedded system debugger must provide access to this state.

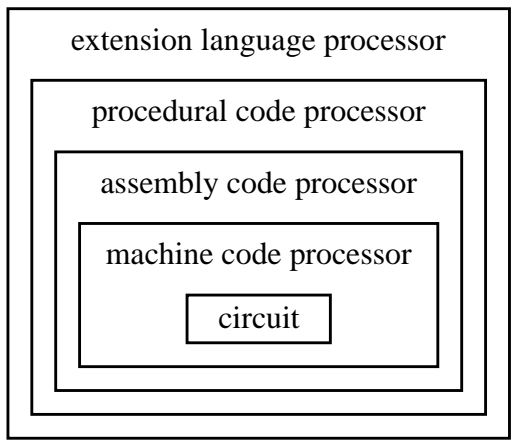

**Figure 2: Layers of Luxdbg virtual machines**

The circuit layer in Luxdbg may be embodied by C++ circuit modeling objects such as memory or register models in a processor simulation model. A simulation model allows users to simulate execution of the instructions and IO operations of an embedded processor system before hardware is available. A simulation model supports debugger inspection of target system internal state and target system timing that may be inaccessible in a real, hardware system because it is hidden within the inside of a complex integrated circuit. The circuit layer may also be embodied by electronic circuits in a hardware processor. Hardware target systems typically provide dedicated debugger access pins and registers (hardware monitors) as well as target-resident debugger access library routines (software monitors) that allow a debugger to monitor and control program execution and target system state. Luxdbg users can interactively read and write circuit registers, memory, and other scalar and vector values using Luxdbg primitives.

The *machine code layer* adds the concepts of an *instruction stream* and a *system clock*. Artifacts such as an instruction pointer (a.k.a. program counter), program memory, hardware interrupts and breakpoints become evident at this level. This layer consists of a programmable processor and the state contained in its circuit layer of abstraction, devoid of symbolic debugging information supplied by compilers, assemblers and linkers. With the machine code layer come Luxdbg primitives for allocating program memory and for determining execution location, primitives for setting and clearing breakpoints at machine program and data addresses, primitives for specifying breakpoint and processor exception handlers, and primitives for resetting, resuming, and interrupting program execution.

System clock-based synchronization of multiple processor cores on a single chip also comes as part of the machine code layer. Timing is an important element of real-time embedded systems. Modern telecommunications embedded systems often employ multiple processors within a single silicon *system on a chip* (SoC). Each SoC houses multiple machine code processors. A SoC simulation model distributes a system-level clock that maintains precise timing relationships among contained processor cores. SoC hardware provides debugger access to execution cycles and instruction counts. Luxdbg can gain access to sub-instruction timing granularity, for cases of debugging interrupt latency or side effects of a visible instruction pipeline, when working with simulation models that represent these precise degrees of timing granularity.

While the previous two layers reside in a target processing system that is being debugged, the remaining layers of Luxdbg reside in the debugger.

The *assembly code layer* adds symbolic interpretation on top of programs running within the machine code abstraction. This is the first layer to relate binary run-time information to build-time source information. Unlike time-sharing systems, embedded systems typically do not carry much symbolic, source code information in the run-time environment. Often the run-time system lacks a loader; programs then reside in ROM. Luxdbg's assembly layer adds a loader, symbol resolver and assembly expression evaluator to each machine being debugged. Primitives at this layer translate program symbolic names to machine code layer memory addresses and contents.

The *procedural code layer* is a more powerful variant of the assembly code layer, adding constructs such as stack frames, data structures and objects that come with source languages such as C and C++. Both the assembly and procedural layers map user-supplied commands that are specified in source code terms to machine addresses and binary values. A C procedural layer, for example, maps a reference to a local C variable to a memory offset from a processor stack pointer in the machine code layer. These layers also map machine code event parameters such as breakpoint addresses up to source code terms such as source file-line number pairs and data structure member names, types and values.

The *extension language layer* is the home of the extension language interpreter. The command set of the extension language layer includes extension language operations and Luxdbg primitives for the inner layers. By supplying primitives from all of these layers to the extension language layer, Luxdbg gives users and scripts access to several perspectives of a target processor. A target embedded processor is simultaneously a collection of embedded circuits, machine registers and memory locations, symbolic data structures and executable functions, and extension language operations.

There are four categories of Luxdbg Tcl primitives.

*Processor management primitives* allow Luxdbg to connect to a set of processors. These primitives include:
- primitives to query the set of available processor simulation models,
- primitives to query hardware debug servers for connected hardware processors,
- primitives to construct a C++ processor model or to reserve a hardware debug connection, and to connect Luxdbg to one or more of these target processors, and
- primitives to disconnect Luxdbg from target processors.

*Process access primitives* allow Luxdbg users to read and write target processor state at all layers of abstraction.
- An *expression evaluator* at each layer reads and writes target processor state, and combines state values within arithmetic expressions. Luxdbg's *fxpr* expression evaluator primitive uses machine code level entities (e.g., registers and memory locations) and assembly level symbols (e.g. labels) to retrieve, compute and store values at these levels. The *ce* expression evaluator evaluates C expressions in the context of the procedural layer, translating symbolic references to machine references. Tcl provides its native *expr* expression evaluator with interactive C-like operations. Tcl can combine results from lower level expression evaluators using expr and other Tcl primitives.
- *Query primitives* allow users to determine the identity of state-bearing entities within each layer of abstraction. Circuit level query allows Luxdbg users and Tcl scripts to determine the identity of registers, circuit signals, and blocks of physical memory in a specific processor. Machine code level query identifies the name of a processor's program counter, a processor's byte order, a processor's native word size, and a processor's program and data memory. (Program and data memory are distinct address arenas in many DSP architectures.) Unlike many debuggers, Luxdbg does not hard code processor-specific details like those listed above, but instead it queries processor simulation models to determine these details at run time. The assembly and procedural levels follow suit by allowing users to query for the identity and type of program symbols.
- Signal logging primitives for simulation models store precise time-value transition tables for pins, registers, and other signals into log files. Time-triggered Tcl procedures can log procedure level information as well.

*Processor control primitives* direct program execution.
- There are primitives to set breakpoints, clear breakpoints, query breakpoints, reset processors, resume execution, stop execution, and to synchronize starting and stopping of multiple processors in a target system.
- Control primitives accept both numeric

addresses for the machine code layer as well as labels, function names, source file-line number pairs and data names from the procedural layer.

*Processor IO primitives* connect processor models and hardware processors to data sources and sinks.

- A loader primitive loads binary values into processor memory regions and loads symbol table information into assembly and procedural debugger layers.
- There are primitives for connecting low-level device IO ports to debugger files and to Tcl callback procedures. IO at the procedural layer usually occurs via library calls, and Luxdbg can insert breakpoints into these calls and redirect library-based IO flow to and from files or Tcl procedures.

Each layer of Luxdbg supports *reflection* [2]. Reflection refers to the ability of client code to interactively inspect the underlying contents of server code, in order to determine unique capabilities of a specific server at run time. Reflection forms the basis for Luxdbg query primitives. One example is a Luxdbg query to determine available processor model types:

luxdbg: pssr models ; # query available models
    16210i 16270i 16410c arm9 ...
luxdbg: ? R ; # query registers in this processor
    {p0 0 s 32} {p1 1000 s 32} {r0 5 u 20} ...

Querying registers yields name-value-typetag-width 4-tuples in this example. As with any Tcl command, results from invocations of debugging primitives are available for use by Tcl scripts. For example this interactive Tcl loop uses the "? R" register query:

foreach reg [? R] { # print register name & width
  puts "reg: [lindex $reg 0] width [lindex $reg 3]"
}

prints this result to Tcl's standard output:

reg: p0 width 32
reg: p1 width 32
reg: r0 width 20
...

Reflection allows Luxdbg to adapt its behavior as well as its user interface to a particular processor at run time. An outer layer can determine and manipulate not only the state of an underlying layer, but through reflection it can also determine the identity of state-bearing entities within that layer. A client of the circuit layer uses reflection to determine the identity and properties of pins, registers and memory blocks within that layer. A client of the machine code layer uses reflection to determine the name of the program counter register, program memory, and byte ordering within a multiple-byte instruction stream. Reflection for assembly and procedural layers exposes symbol names, types and addresses, stack frame conventions and source file identities. Reflection for an extension language such as Tcl provides information on the state of the interpreter, for example:

info commands ; # queries Tcl command set
    next logsigs down resume ...

Luxdbg uses reflection to adapt its operations to each target processor and program. Such flexibility is essential in debugging multiprocessor systems that have come to include an array of heterogeneous processor types and programming languages.

Luxdbg's *expression evaluators* rely heavily on reflection. Luxdbg's *fxpr*, the machine code and assembly language expression evaluator, uses fixed syntax that is compatible with *expr*, Tcl's built-in arithmetic expression evaluator. Fxpr operands, however, come from machine code entities (e.g., registers and memory) and assembly language symbols that fxpr identifies and reads or writes via reflection. Fxpr can thus adapt itself to different processor architectures and target programs by querying their reflection interfaces for operands available to fxpr. Likewise, Luxdbg's C expression evaluator, *ce*, queries target programs for available types, variables, functions, etc. The extension language layer has access to all of this reflection and the primitives built using it.

Each processor object has a textual *instance name* in Luxdbg. A user can invoke the *processor new* command to create a new processor simulation model or to connect to a remote processor or process, and to associate this processor with an instance name such as "dsp1" or "controller" or any user-selected name. Thereafter the user can use an instance name as a prefix to debugging commands. A Luxdbg primitive uses an instance name to set the *current processor* reference in C++. Commands nested within the dynamic scope of an instance name go to that processor object. For example, the Tcl command "p2 fxpr r0 = [p1 r3] * 2" retrieves the value of register r3 from processor p1, multiplies it by 2, and stores the result in register r0 of p2. Fxpr determines the semantics of its arithmetic operators by consulting its target machine code processor. Fixed-point DSPs supply fixed-point semantics while microcontrollers supply mixed floating point / integer semantics similar to C expressions.

A Luxdbg Tcl expression can defer specification of a target processor. Suppose, for example, Tcl procedure *logRegisters* were written, without a processor prefix, to write name-value pairs for all processor registers to standard output. Suppose further that Tcl procedure *logAll* were written, again without a processor prefix, to invoke *logRegisters* along with some other log procedures. Now suppose that Luxdbg is connected to a DSP instance that a user has named "dsp5," and the user invokes "dsp5 logAll" in an interactive command or from within a Tcl script. Procedure *logRegisters* works with Luxdbg's *current processor*, in this case the processor named *dsp5*. Primitives such as fxpr query dsp5 for state-bearing entities and their contents. The Tcl procedure logRegisters, while not a primitive, uses primitives to determine the identity and contents of registers within the current processor, without any hard-coded knowledge of that processor's registers. A later invocation of "controller2 logAll" for microcontroller "controller2" would perform similarly for controller2, which might be an entirely different sort of processor. As long as logRegisters and logAll are written to use the reflection interface and target-neutral primitive commands, and to avoid making target processor assumptions, they can work for a variety of processors, selected by a user or script at run time.

So far we have described the basic extension language, virtual processor and reflection machinery that provides the basis for Luxdbg's debugging power. The next section catalogs a number of ways in which Luxdbg developers, field support staff and users can employ this machinery in extending the capabilities of Luxdbg as well as its target processors.

## 4. Luxdbg avenues of extension

### 4.1 Command and callback path mechanisms

The *command path* of Figure 1 initiates activity in Luxdbg. All interactions start out when a user, a Tcl script or an auxiliary tool issues a set of textual commands to Tcl. Tcl, in turn, examines and modifies target processor state. Tcl sets breakpoints via breakpoint primitives; it causes target processor execution by calling the *resume* primitive. Tcl again interacts with processor state upon processor arrival at the next breakpoint. These are all examples of command path control mechanisms. Upper, client layers direct control.

*Callback path* control comes about when a target processor object reacts to an event in the target processing environment by invoking a callback procedure. A callback may be a Tcl expression set by a user, script or auxiliary tool. A callback may also be a C++ method, built into Luxdbg, that reacts to target events. A target event causes its processor to stop at a breakpoint, and a callback can interact with the stopped processor as well as with any other Luxdbg processor that is at a breakpoint. Luxdbg sets the default *current processor* to the triggering processor during a callback. A single processor may provide multiple target events, thereby triggering multiple callbacks, during a single interaction with the debugger. If all events connect to callbacks, and if all callbacks invoke *resume*, then upon completion of all callbacks, the target processor resumes execution as though no breakpoint had occurred. Callbacks form the basis for *conditional breakpoints*.

Command and callback operations have one important difference, part of which we have just seen. Invoking *resume* as a command operation causes a halted processor to begin execution. Invoking *resume* as part of a callback operation schedules that processor for resumption upon completion of all callbacks. A single callback invocation that does not invoke *resume* causes a break to the outer, command-invoking user, script or auxiliary tool upon callback completion. An outer, command *resume* invocation can start any processor, while a callback *resume* can schedule only the processor that triggered the callback.

The major net effect of callbacks is the extension of target processor primitives at any of the layers of Figure 2. From the perspective of a command *resume* invocation, execution of a callback can appear as a seamless part of target processor execution. All of the examples of this section use callbacks to extend processor capabilities.

## 4.2 Conditional debugging and assertions

Callbacks support incremental extension of conventional debugging activities, from providing small helper functions that eliminate manual activity, to providing sophisticated assertion checking mechanisms. Conditional breakpoints provide one example. Suppose a target C function exhibits an error after many invocations of that function. A simple breakpoint that stops on every function invocation becomes annoying because of the number of times that the debugger user must manually resume execution. Some debuggers provide breakpoint counters that a user can set in order to skip over breakpoints, but these debuggers do not provide means to determine the counter value in the first place. They also do not provide a means for sampling the triggered breakpoints, for example stopping on every tenth breakpoint. A user could insert extra debugging code into a target function to keep track of invocation counts, but this insertion involves time-consuming compilation, and it shifts the target program address space, possibly masking the error being debugged.

Listing 1 shows Tcl procedure *setCount* that counts the number of invocations of target C function *targetFunc*. Assume that the user has issued the command "stop in targetFunc setCount," which sets a breakpoint in targetFunc and associates setCount as the callback operation to pass to Tcl. Every target invocation of targetFunc triggers the breakpoint, which stops the target and invokes setCount via the callback path. Tcl interprets setCount, incrementing its counter and resuming target execution. Eventually the target error occurs, causing a breakpoint for which the user has not provided a callback, so control returns to the user. Now the user can interactively retrieve the counter value by issuing the "set setCounter" command to Tcl.

```
set setCounter 0 ; # initialization
proc setCount args {
    global setCounter
    incr setCounter ; # bump counter
    resume ; # continue target processor execution
}
```

**Listing 1: Recording a target invocation count**

With a setCounter value in hand (e.g., 1233), the user could rewrite the "resume" line of procedure setCount to be a conditional: "if {$setCounter < 1230} resume." Since Tcl is interpreted, the user can rewrite setCount interactively and use its new definition immediately to stop function targetFunc in the desired invocation by restarting the target application.

Now setCount is the simple counted breakpoint feature built into some debuggers. Sampling every tenth breakpoint could be achieved by using Tcl's modulo ("%") operator:

    if {($setCounter % 10) != 0} resume

In general, any callback expression of the form

    if {predicate} resume

supports conditional breakpoints, where "predicate" could be the invocation of a Tcl procedure. Execution breaks to the user when a predicate is false. Predicate testing can occur at any level of abstraction of Figure 2. For example

    if {[fxpr r0 != endlocation]} resume

uses fxpr to compare machine register *r0* to the address of assembly label *endlocation*. Predicate

    if {[ce head_of_list != NULL]} resume

uses the C expression evaluator to compare C variable *head_of_list* to the NULL value. Predicate

    if {[info commands $pname] == ""} resume

uses Tcl's *info* command to test for the existence of a command whose name is contained in variable *pname*.

This technique is much more powerful than standard conditional breakpoints, because the latter allow tests only on the target program. A Luxdbg user can extend the test space available to predicates. The test space could be the saved results of previous target runs, allowing breakpoint callbacks to monitor regression test results.

The test space may also be a set of assertions that a programmer expects to see upon entry or exit of a target function. By setting breakpoints at function invocation and exit points, and by coding assertions as Tcl predicate expressions, the programmer can check assertions without modifying target code, halting execution only when an assertion is violated. Since the assertion checker is a piece of Tcl code, it can print a message identifying the cause of failure to the user (e.g., print the assertion expression as a string). When control returns to the user, source code display identifies the location at which the assertion failed.

## 4.3 Extended input-output

Callbacks support input-output extension for all of the layers of Figure 2. At the lowest layer, a processor simulation model can include *peripheral models* that take the place of actual IO hardware. Each peripheral model contains simulated registers and memory that represent their hardware counterparts. In addition, each peripheral model contains a callback hook that a user can connect to a Tcl procedure. Suppose a target processor executes the machine code instruction "r0 = pio1," where *pio1* is a parallel IO port register. In real hardware the processor would latch the value on the input pins of pio1 into a register and transfer that value to register *r0*. In a Luxdbg simulation where the user has connected pio1 to a Tcl procedure by using the *srcfn* command (to "source" an input from a function):
> srcfn pio1 myInputProc

Luxdbg invokes a callback to the Tcl procedure *myInputProc*. The procedure can retrieve a value for pio1 from many places — a file, the user, an output port from a different processor instance, or from a random number generator. The return value from the Tcl procedure finds its way into simulated register pio1, and the simulation continues without interruption.

Reversing the machine code instruction to its output equivalent, "pio1 = r0," copies r0 to a parallel output port in real hardware. In a Luxdbg simulation where the user has connected pio1 to a Tcl procedure by using the *sinkfn* command (to "sink" an output to a function):
> sinkfn pio1 myOutputProc

Luxdbg again invokes a callback. This time the callback procedure *myOutputProc* receives the output value as a parameter, and it can write the value to a file, user, or input port of another processor instance.

Input-output redirection is not restricted to simulation models. Both models and real hardware can use memory access breakpoints (a.k.a. data breakpoints) to simulate *memory-mapped IO*. A read or write for a memory location with a breakpoint results in a callback that simulates memory-mapped IO. The *stop* command can specify breakpoints at the machine code, assembly or procedural levels of abstraction, supporting memory-mapped IO simulation at any of these layers. For example, if a C program contains a global variable called outport:
> int outport ;

that the C program uses to simulate output, Luxdbg can set a data-write breakpoint on this variable:
> stop outport -write outportHandler

and Tcl procedure *outportHandler* can use *ce* to retrieve outport for writing to a file:
> puts $logfile [ce outport %d]

Comparable machine level memory-mapped IO on address 0x2000 would use "stop 0x2000 -write handler" to set the breakpoint and "fxpr *0x2000" within procedure *handler* to retrieve its value.

Our trivial examples only hint at the possibilities. The appendix describes a simulation model for an entire application-specific IC block (ASIC) written as a set of Tcl functions. A multiple-processor chip could use such a hardware ASIC for inter-processor communications. We can simulate this interaction in Luxdbg by running real, discrete hardware processors, routing their communications through the Tcl ASIC model via breakpoint callbacks.

Luxdbg supports IO redirection at higher levels of processor abstraction. A target embedded system might not support textual IO, but C *printf* statements remain a popular method of monitoring and debugging program execution, even in the presence of powerful debugging utilities. Because of the usefulness of streaming textual IO during software development, Luxdbg supports *semi-hosted libraries* for many of the standard C IO system calls and library functions. Listing 2 shows the very minimal portion of printf implemented for a target DSP16000. This assembly code links directly into an embedded application running in hardware or in a processor simulation model. It works by putting the printf function identifier into the processor's *i* register, then invoking the *icall 0* system call instruction. This system call traps the processor or operating system process, which then runs a breakpoint handler, and a breakpoint event arrives at Luxdbg.

Listing 3 shows the beginning and end of the event callback for the breakpoint-based printf. It reads the printf format string from the top of the DSP16000 stack memory, iterates through a loop that reads value parameters and concatenates an output string (details not shown), sends this string to the debugger's standard output, and finally returns the number of characters printed back to the target environment in register a0.

```
#include "shlib.h"

.rsect ".text"
.align
function ___printf(
    // IN: *(sp) = format string
    // IN: *(sp+N) = other arguments (if used)
    // OUT: a0= output length on success, else -1
) {
    i = _SP_FUN_CODE(_SHLIB_PRINTF)
    icall 0
    return
}
```

**Listing 2: Printf hook in a DSP16000 target**

```
proc handle_printf { } {
   set top [sp @rd]; a0=-1
   set sFormat [readstring [list ymem \
    [readlong "ymem $top"]]]
   incr top 2; set skipme 1; set formatargs ""
   foreach arg [split $sFormat %] {
       ... (Implementation not shown)
   }
   set output [eval format \$sFormat $formatargs]
   send_to_stdout $output
   a0=[string length $output]
 }
```

**Listing 3: Printf implementation snippet in Tcl**

Input-output handling at the client layer of Figure 1 takes the form of Tcl/Tk GUI extensions. Luxdbg gives users access to the full power of the Tcl/Tk graphical tool kit [4], with which they can create custom graphical widgets and bit-mapped graphical canvases. As with the other levels of IO redirection, users can connect event callbacks to GUI extensions, updating their graphical creations with processor information as changes occur.

### 4.4 Multiple processor scheduling

Luxdbg becomes involved in processor scheduling when it is used to simulate communications within a multiprocessor system. Luxdbg loads one or more frames of input data to the first processor in a pipeline or other multiprocessor topology, then uses *resume* and breakpoint callbacks to run that processor until it fills an output buffer. Filling an output buffer triggers a breakpoint, and a breakpoint callback procedure copies the output buffer to the input buffer of a receiving processor. A Tcl level scheduler can then *resume* the receiving processor's execution, causing it to contribute its part to system data flow. This model of multiprocessor communication forms the basis of the demo described in the appendix.

Processor input-output handling works using the callback path of Figure 1. Processor scheduling takes us back to the top-down command path. Recall from Section 4.1 that invoking *resume* from above causes processor execution, while invoking *resume* from within a callback avoids taking an interactive breakpoint. Avoidance of callback *resume* is the key to Tcl schedulers. A breakpoint that has no Tcl callback, or a callback without *resume*, causes processor execution (initiated via *resume*) to stop, returning control to the outer *resume*'s caller. This outer *resume* may have come from a user, or it may have come from a Tcl scheduler script.

Listing 4 shows a simple round-robin, cooperative scheduler for a list of processors passed as a parameter to this Tcl procedure. Each processor in the list runs until it hits a breakpoint, and the scheduler iterates over the list *N* times. Presumably the breakpoint is triggered on the completion of some task, e.g., completion of processing a data flow. A scheduler with uncooperative processors could substitute *stepi $count* for *resume*; the *stepi* command steps a target processor some number of machine level instructions; it is guaranteed to return.

```
proc scheduler {processors N}{ # command path
   for {set i 0} {$i < $N} {incr i} {
     foreach p $processors {
         $p resume
     }
   }
 }
```

**Listing 4: Round-robin, cooperative scheduler**

Listing 5 gives a scheduler variation, this time a scheduler driven by processor output events. The scheduler starts out using parameter *startproc* to identify the processor to start, then it consults a global Tcl variable *nextproc*. Output breakpoint callback procedure *output_callback* is coded to write its output value to a neighboring processor's input port — obtained via associative array *neighbor*, set up by the scheduler's Tcl script, which is indexed on the current processor name obtained via primitive *processor name* — and then pass that neighbor's name to the scheduler for execution via nextproc. The lack of *resume* in the callback guarantees return of control to the command path scheduler. Scheduling terminates when the last processor in the sequence is reached, signified by a blank slot ("") in the neighbor table.

```
proc scheduler {startproc}{ # command path
    global nextproc
    $startproc resume
    while {$nextproc != ""} {
        # terminate on empty processor name
        $nextproc resume
    }
}
# output callback is attached to each processor's
# output breakpoint
proc output_callback args { # callback path
    global nextproc neighbor
    set nextproc $neighbor([processor name])
    if {$nextproc != ""} {
        # copy my outport to neighbor's inport
        $nextproc fxpr inport = [fxpr outport]
    }
    # no resume means break to scheduler
}
```

**Listing 5: IO event-driven scheduler**

Luxdbg originally provided only the blocking *resume* and *stepi* commands for execution resumption, allowing only one processor or multi-processor chip to execute at one time. Serial scheduling was the only possibility. Luxdbg now includes the ability to run target processors in the background without blocking the debugger via the *resume &* command, named after the UNIX use of *&* for background execution. We are adding a *wait* command that will block until specified background processors have stopped at breakpoints. The combination of *resume &* and *wait* gives Luxdbg logical *fork* and *join* operations for concurrent target processors and processes. With these we will be able to write concurrent schedulers that interleave target processor execution, blocking the debugger until it is safe for it to act.

### 4.5 Exception logging and testing

Luxdbg treats exception events in a target processor similarly to the way it treats breakpoints. Each exception has a unique identifier to which a Luxdbg user or script can attach a callback. Exceptions come in four levels of severity — note, warning, error and fatal. Default behavior for notes and warnings is for Luxdbg to print a message and issue *resume*, while default behavior for errors and fatal errors is to print a message and stop the processor. Callbacks can be attached to any exception type, and any non-fatal exception callback can successfully invoke *resume*. Users can use this mechanism to shut off unwanted notes and warnings, to log specific exceptions, and to perform processor exception handling in Tcl similar to breakpoint handling already discussed.

Among other uses, an extension language such as Tcl provides an ideal basis for regression test machinery. Tcl has a complete set of file manipulation operations, making it possible to set up tests and compare results from files of text tables. Tcl provides a *catch* instruction that allows Tcl-level handling of failed instructions, so it is possible to set up degenerate conditions in a test script and test for proper Luxdbg reactions without crashing the script if Luxdbg operations fail. Luxdbg testing uses all of the extension language machinery discussed above within its regression test suite.

### 5. Related work

Several other debuggers have used debugging languages or extension languages. The *deet* debugger for C programs [6] comes closest to Luxdbg in its use of Tcl as the user interaction language, its use of a user-extensible Tk GUI, and its application of Tcl for conditional breakpoints and testing. Deet builds atop cdb, an earlier, machine-independent debugger for C programs [7]. Luxdbg and deet differ in their use of Tcl primarily with respect to Luxdbg's user association of arbitrary Tcl expressions with breakpoints and exceptions for callback interactions. Deet supports conditional breakpoint expressions that it evaluates using Tcl procedures, but deet appears not to have Luxdbg's more general target processor event handling machinery. Much of this Luxdbg machinery finds application beyond debugging, for example in simulation, testing and prototyping, but event-driven callbacks are also useful for automating reactions to bugs.

SmartGDB is another Tcl-based debugger [8]. In addition to adding GDB commands as primitives to a Tcl interpreter, SmartGDB provides the ability to associate a Tcl procedure with a breakpoint event. SmartGDB does not distinguish between the command path and callback path of Figure 1. Invoking the GDB *continue* primitive from with a breakpoint procedure causes immediate resumption of processor execution, resulting in loss of breakpoint context as well as loss of any concurrently triggered breakpoint procedures. Because of the lack of distinction between command and callback invocations of the Tcl interpreter, SmartGDB does not support seamless extension of debugger primitives via resume-bearing callbacks.

*DUEL* [9] and *ACID* [10] are two debuggers with embedded, special-purpose debugging languages. These languages are not general-purpose scripting languages such as Tcl, but rather are dedicated to debugging specific types of target systems and languages. As with deet, both use top-down, command interactions from the debugger to a target system.

Generalized path expressions [11] and the Formal Annotation Language (FORMAN) [12] are two query-oriented debugger languages that operate on program execution traces. Path expressions provide a special purpose notation for querying program execution paths and for specifying ordering and other constraints. FORMAN uses target processing events and a general event grammar to provide a language for automation of debugging, assertion checking and profiling. Both approaches operate at a higher level of abstraction than Luxdbg, providing mechanisms for searching and testing the execution history of a running process.

In the issue of language selection, there is a trade-off between language applicability and language accessibility to a large programming population. Use of a special purpose language requires learning yet another programming language that is dedicated to a single stage of software development, debugging. Use of Tcl, on the other hand, transfers into areas beyond debugging, such as testing, user interface design, and other applications. Popular extension languages come with large utility libraries, commercial support, and a large user community from which to draw information. With modest efforts a user can build or collect Tcl procedures for Luxdbg that achieve some of the benefits of the debugger-specific languages. We have found that field application engineers and expert users of Luxdbg have built sophisticated, application-oriented debugging infrastructure using Tcl that easily puts Luxdbg on par with dedicated debugging languages with respect to expressiveness. Tcl could provide a base for building automated debugging extensions similar to those of dedicated debugging languages.

The Coca debugger has used a more widely accepted, interactive language — PROLOG — as the basis for a powerful engine that searches program trace information during target program execution [13]. Coca uses PROLOG's backtracking on failure mechanism very effectively to allow users to explore a set of program execution states that satisfy query constraints. Coca avoids the storage-intensive performance problems of some relational debugging systems by using breakpoints and target system-resident program event extraction and analysis modules to perform its queries at target run time. With the ready availability of PROLOG interpreters, it would be possible to replace Tcl with PROLOG in a research version of Luxdbg that supports run-time selection of the extension language [14]. A novel requirement of Coca is extension of the symbol table information of a target program with annotations concerning various target programming constructs such as the location of *for* loops and *if* conditional constructs for language construct-driven specification of queries and corresponding placement of breakpoints.

The *ldb* debugger uses PostScript as an extension language for communicating target program symbol table information among the compiler, target application and debugger [15]. It does not bring the extension language out to the debugger user.

Luxdbg differs from all of these related projects in several regards. Luxdbg uses processor / process instance names, as Tcl command prefixes, to delimit the dynamic scope of debugging commands. Luxdbg provides debugging at the multiple layers of abstraction of Figure 2. It includes dedicated expression evaluators *fxpr*, *ce* and *expr* to delimit the scope of symbols within an expression. DUEL and ACID work directly in target system scope, at a single level of abstraction, while deet employs a single level of namespace indirection for C variables. The other debuggers cited use explicit scope to access variables at a single level of target language abstraction. Consequently Luxdbg could be viewed as a series of debuggers, for different levels of processor abstraction, with an expression evaluator for each. Commands that interact with more than one level of abstraction, such as *stop* for setting breakpoints at machine, assembly or procedural levels of abstraction, include command syntax for disambiguating context. Commands infer a default context of either a procedural language or assembly / machine code, based on a processor's current breakpoint address and corresponding program scope.

Many of the features of Luxdbg are orthogonal to the issue of extension languages. Virtual processor layers, reflection, multiple processor debugging and heterogeneous processor debugging are examples. Yet this orthogonality is a property that gives an extension language considerably more power in Luxdbg. An extension language user has access to all of these orthogonal capabilities, and a user can combine these capabilities via extension language scripts to create powerful custom debugging features.

## 6. Conclusions and directions

Luxdbg is being enhanced, supported and deployed by a team of engineers in Lucent Microelectronics. Its Tcl extension language features are very successful. All of the aspects of Tcl discussed in this paper — its use in mixed C-assembly-machine code-circuit debugging, its use in customizing embedded input-output operations and exception callbacks, its use in processor scheduling, its use in testing, its use in constructing custom development environments, and its use in extensible GUI construction — find practical application in the field. Luxdbg's application of Tcl remains one of the strongest aspects of this debugger.

Ongoing work includes investigation of Java for upper debugger layers in order to take advantage of Java networking, dynamic loading, reflection and class library infrastructure. Luxdbg has used Java to parameterize the extension language, so that languages such as Python can be loaded at run-time in the place of Tcl [14].

Distributed debugging supported by distributed extension language interpreters is an area for research [16]. Passing expressions across the Internet to remote debugging sites is easily achieved with interpreted extension languages. The *send* command provides the basis for remote interpretation in Tcl [4]. By combining the debugging primitives of Luxdbg with communications primitives such as send, extension languages provide a powerful base for distributed debugging research and application.

## 7. Acknowledgments

# 8. Appendix — Demonstration: Exploiting Extension Language Capabilities in Hardware / Software Co-design and Co-verification

Cathy Moeller, Dale Parson, Bryan Schlieder, Jay Wilshire
{cbmoeller,dparson,bryanschlieder,wilshire}@lucent.com

**Abstract**

*Proper design of an embedded system requires the ability to analyze multiple implementation options to determine what parts of the system are best implemented in hardware and what parts are best implemented in software. Construction of a hardware prototype is both expensive and time consuming. In this demo we use the Tcl extension language to explore large parts of an embedded system design space, to debug and to optimize software before hardware is available, and to utilize existing hardware to accelerate the design process. The demo illustrates how abstracting differences between simulation and hardware emulation can minimize code size and maximize test software reuse.*

## 8.1 Introduction

The Lucent Technologies LUxWORKS [17] debugger, Luxdbg, uses Tcl/Tk [18] as its interactive command language and user interface construction library. The Luxdbg Graphical User Interface (GUI) is written entirely in standard Tcl/Tk and communicates with a modified Tcl command interpreter. The modified interpreter adds debugging commands to support debugging of both processor simulation models and target hardware systems that can be controlled through an emulation interface: either classic pin-based in-circuit emulation or serial IEEE 1149.1 test port (a.k.a. "JTAG") [19] access.

Luxdbg's rich set of breakpoint capabilities, device state observation and control commands and graphic extensions enables complex interactions with either the device model, target hardware, or both. The Tcl extension language includes support for separately loadable packages and unique namespaces so that users can easily write application-specific Luxdbg extensions.

Tcl serves as an interactive command language and as a programmable scripting language for initiating LUxWORKS debugging actions. In addition, Tcl serves as an event handling language for processor events such as breakpoints and exceptions. A user can associate a Tcl expression with a processor breakpoint or exception. When the associated processor event occurs during execution, the processor stops and Luxdbg invokes the corresponding Tcl expression as a callback. This Tcl callback can read, write, and copy processor state among the set of stopped processors. The callback can also resume execution of its stopped processor, or it can force a break to the user or outer Tcl script.

Tcl/Tk allows users to customize the Luxdbg interface by creating and modifying menus, buttons, and windows and attaching them to user-defined scripts. The demonstration shows the flexibility and power of Tcl/Tk scripting by extending the LUxWORKS debugger and its target processor systems. We use Tcl/Tk to:

- create an abstract model of an interprocessor communication channel in the form of a bidirectional shared-memory mailbox that is used with both simulation models and/or target hardware to study the effect of various mailbox architectures on key system performance factors such as latency (polling vs. interrupts), throughput, processor communication overhead, and application software complexity,

- produce a reusable library of debugging and profiling scripts to verify the correctness and performance of the system,

- provide high-level visibility into the workings (or non-workings) of various system configurations through a custom graphical interface built from freely available Tcl/Tk graphics packages [20].

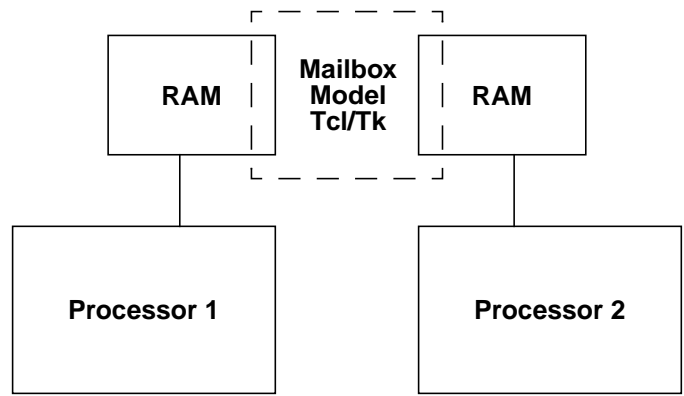

**Figure 3: Demonstration Components**

### 8.2 Approach

Figure 3 shows the topology of the demonstration. The two processor building blocks are identical Luxdbg processor instances with access to conventional external RAM. Identical instances simplify implementation, but the demonstration easily extends to heterogeneous processor instances. The Mailbox Model is a simulation model implemented as a set of Tcl procedures that use breakpoint callbacks to gain control, and memory-to-memory transfers to give the appearance of a shared-memory space.

The processor instances are either device simulation models or actual devices on a circuit board controlled by Luxdbg that have access to external memory. The choice of simulated models, hardware targets or a combination is made at run time.

Data buffers in conventional memory of the two processors hold message data and represent the control and status registers of the simulated shared-memory mailbox circuitry.

Data address breakpoints trap writes to the simulated mailbox control register of each device. When the sender issues a command, the Tcl callback associated with the breakpoint simulates mailbox hardware by copying data from the sender to the receiver's memory, updating the receiver's status register, and optionally redirecting the receiver to an interrupt service routine.

Test software running on each processor exercises driver software tailored to specific simulated mail-box hardware. Both mail-box hardware and driver-software defects are detected by running automated simulations and/or hardware emulations. A library of test procedures written in Tcl helps the user explore the test space. Users select hardware/software configurations and run tests that:

- send increasingly large messages in both directions simultaneously,
- attempt to starve the receiver in a polled environment,
- measure throughput, latency and processing overhead.

A graphical interface controls and displays test parameters and displays test results.

### 8.3 Demo Details

The target hardware is a Lucent DSP16210 demonstration board with 2 DSP16210 processors, one with 64 Kilobytes and one with 256 Kilobytes of external data memory. The only physical connection between the processors used in the demo is a shared JTAG serial test bus. In hardware mode, memory transfers pass data from one processor to the debugger-resident Tcl Mailbox Model over JTAG, and then down to the other processor over JTAG.

The demo can also use simulation models of the DSP16210 processor, each including the necessary memory simulation models. The debugger interacts with each model instance via "software probes" that are C++ objects that read and write processor state-bearing objects. These software probes take the place of processor JTAG ports used for hardware debugging.

Simulation within Luxdbg is single threaded, so dual-processor simulation requires a scheduler written in Tcl to pass control between two processor model instances. An extremely simple scheduler that advances each model by one instruction is implemented in six lines of Tcl:

```
# Simulation mode step-based scheduler,
# default step count is 1.
proc advanceSim { p1 p2 { count 1 } } {
    for { set ii 0 } { $ii < $count } { incr ii } {
        $p1 stepi
        $p2 stepi
    }
}
```

P1 and p2 are Tcl variables that contain the processor instance names; count is the number of steps to advance. Even this simplistic advance procedure shows the power of Tcl: formal parameters p1 and p2 provide reusability and the default value 1 for count helps makes the syntax compact.

In hardware mode, the advance procedure could also be implemented with a step command. However, each invocation of debugging primitive "stepi" entails flushing debugger state to the processor, stepping the target instruction, and then retrieving state to the debugger. Each transfer of hardware state involves communication overhead not entailed by simulation mode, since a simulated model and its probes reside in the debugger process. A more efficient approach is to run each hardware processor until its reaches a well-defined control point (e.g., write to a mailbox, or run until cycle count is exhausted), and then break to the debugger. Tcl procedure *advanceHdw* executes its target processors for *count* machine cycles; *count* gives a time slice for each processor.

```
# Hardware mode cycle-based time-slice scheduler
# Run processor p1 then p2 for count cycles
proc advanceHdw { p1 p2 { count 1 } } {
    $p1 runProcessorForCycles $count
    $p2 runProcessorForCycles $count
}
```

The demo's advance procedure uses both *advanceSim* and *advanceHdw* and also handles mixed modes with one device in hardware mode and one in simulation mode.

Procedure *runProcessorForCycles* uses low-level, hardware-specific registers *cycles* and *cyclec* for counting and controlling execution cycles, and pseudo-registers *tryCycles* and *elapsedCycles* for tracking cycles. Users can add pseudo-registers at run time. Some low-level registers such as *cyclec* are inaccessible to typical users, but "configure -hiddenregisters on" adds hidden registers to a processor's reflection API, rendering these registers visible in Tcl. *RunProcessorForCycles* deals with the fact that cycle-counting hardware breakpoints have not been integrated into the Luxdbg *at* command that sets temporal breakpoints. *RunProcessorForCycles* sets hardware cycle counting registers directly, and it installs procedure *expired* as a callback handler for exception UNHANDLED_BP. A target processor throws UNHANDLED_BP when any target breakpoint occurs that was not explicitly set from the debugger via *stop* or *at*. In this case the cycle counter causes an UNHANDLED_BP exception because *runProcessorForCycles* bypasses *at* and sets *cyclec* and *cycles* directly.

```
# advance the state of current hardware processor
# - instance specific
set UNHANDLED_BP 5067
proc runProcessorForCycles { docycles } {
    global UNHANDLED_BP
    # register the breakpoint exception callback
    except $UNHANDLED_BP expired
    configure -hiddenregisters on
    # request DEBUGMODE on cycle countdown
    cyclec = cyclec | 4
    tryCycles = $docycles ;# record requested stride
    elapsedCycles = cycles ;# save
    # count up - rollover causes DEBUGMODE brkpoint
    cycles = 0 - tryCycles
    resume            # wait for callback
}
```

Expressions such as "cyclec = cyclec | 4" in *runProcessorForCycles* are implicit invocations of fxpr. Luxdbg intercepts Tcl's *unknown* command to determine whether an unknown command is a valid fxpr expression, and if so then the interceptor invokes fxpr.

Callback procedure *expired* tests to determine whether the thrown exception matches the desired cycle count breakpoint exception, and if it does, *expired* then clears the breakpoint bit from the *cyclec* register, restores *cycles* from being a cycle breakpoint trigger to a cycle counter, and removes the hidden registers from the Tcl reflection API.

```
# procedure to handle cycle counter expiration
# - instance specific
set CYCCNT_EXPIRED 2415919104
proc expired { errnum severity errstr } {
    global CYCCNT_EXPIRED
    # make sure this is our DEBUGMODE brkpoint
    regexp {^.*handle ([0-9]+) *$} "$errstr" match Value
    if { $Value == $CYCCNT_EXPIRED } {
        # clear DEBUGMODE request
        cyclec = cyclec & 0xfb
        # restore cycles, adjust for instruction boundary
        cycles = elapsedCycles + tryCycles + cycles
        configure -hiddenregisters off
    }
}
```

Comparing *advanceSim* and *advanceHdw* highlights the fact that *advanceSim* is implemented directly in terms of the machine code layer of Figure 2, while *advanceHdw* digs down into the circuit layer by calling *runProcessorForCycles* and *expired*. Each *advanceSim* invocation of "$p1 stepi" and "$p2 stepi" causes execution of one target machine code-level instruction. Each *advanceHdw* invocation of *runProcessorForCycles*, on the other hand, runs a processor for a real-time slice by exposing and manipulating circuit-level registers that cause circuit-level effects. This particular circuit-level effect, a time-out breakpoint, is not even integrated into Luxdbg's official breakpoint machinery via the *at* command. But by handling an unknown breakpoint exception in Tcl procedure *expired*, which extracts cycle count status from circuit-level registers, a Luxdbg user is able to compensate for lack of integration of cycle breakpoints into *at*. Tcl procedures *runProcessorForCycles* and *expired* encapsulate circuit-level details within themselves, providing Tcl-interpreted support for the machine code-level semantics of *advanceHdw*.

With *advanceSim* and *advanceHdw* we have the means to interleave the execution of our two processors, and we now go to the matter of communications via the mailbox. Here we give only the most general implementation of a mailbox mechanism, devoid of modeled circuitry and precise timing. The actual demo looks at alternative, circuit-level implementations of the mailbox.

Suppose the program on each of our target processors uses assembly level variables to simulate incoming and outgoing buffers in a mailbox. Memory arrays *sendBuffer* and *recvBuffer* are outgoing and incoming memory buffers. Assembly locations *sendLock* and *recvLock* hold booleans that control their buffers, and *sendCount* and *recvCount* hold the length of a buffered message stream. These assembly-specified storage locations simulate memory-mapped peripherals in a real device.

A processor uses a test-and-set instruction on *sendLock* to gain control of the send buffer; a non-zero value for *sendLock* signifies an unlocked buffer, while a 0 signifies a lock held by the buffer's processor or by the mailbox. A processor adds message data to the stream in *sendBuffer* only when the processor holds *sendLock*. Likewise, a processor drains message data from the stream in *recvBuffer* only when it holds *recvLock*. Conversely, the mailbox drains a *sendBuffer* when it holds the sending processor's *sendLock* and it fills a *recvBuffer* when it holds the receiving processor's *recvLock*.

We can take advantage of the default, single-threaded control mechanism of Luxdbg, and its use of breakpoint-driven control, to build a simple test driver that uses polling interaction between the mailbox and each processor.

```
proc advance {p1 p2 {count 1} {isSim 1}} {
    # run the processors a time slice each
    if {$isSim} {    # simulation mode
        advanceSim $p1 $p2 $count
    } else {         # hardware mode
        advanceHdw $p1 $p2 $count
    }
    # now interleave mailbox polling
    pollMailbox $p1 $p2
    pollMailbox $p2 $p1
}
```

Procedure *pollMailbox* inspects the locks and counts for its sender's and receiver's buffers to ensure that a message is ready and that there is room for it.

```
proc pollMailbox {sender receiver} {
    # Test-and-set is atomic because sender and receiver
    # are stopped; fine-grain simulation will require
    # more precise modeling of test-and-test.
    # fxpr "@rd" returns result to Tcl in decimal
    if {[$sender *sendLock != 0\
            && *sendCount > 0 @rd]} {
        if {[$receiver *recvLock != 0 \
                && *recvCount == 0 @rd]} {
            # memory vector copy via fxpr:
            $receiver fxpr recvBuffer(0) = \
                $sender\::sendBuffer(0:*sendCount-1)
            $receiver *recvCount =
                [$sender *sendCount]
            $sender *sendCount = 0
        }
    }
}
```

Inter-processor memory transfer uses fxpr's ability to copy a block of memory contents between processors without bringing these values out to Tcl. Tcl stores its return values as strings, and copying memory vectors one-element-at-a-time in a Tcl loop would entail binary-to-ASCII conversion when reading a value and ASCII-to-binary conversion when writing it. Fxpr vector assignment copies binary values directly.

This first-cut prototype of *pollMailbox* makes several simplifying assumptions:
- Message transfer is all-or-none. A message transfers only when its receiver's buffer is completely empty.
- Receiving capacity is sufficient. There is no test on the size of the message.
- Simulation is single threaded. The tests of sendLock and recvLock never set these locks to 0 as part of test-and-set.
- Timing granularity is determined by driver procedure *advance*, and *pollMailbox* can transfer the entire message, and release its locks, within the allotted time.
- Message transfer is implemented using polling.

The demo progresses in a manner similar to real prototyping, by taking an overly simple solution and elaborating it into a realistic one. Progressive variations build finer timing resolution, overflow detection, a robust test-and-set operation and interrupt-driven control into the mailbox. Some Luxdbg simulation models support sub-instruction, *phase-accurate timing*. Such a model can return control to Tcl at every transition of the simulated hardware clock signal that drives the model's state transitions. Using such a model, it would be possible to simulate precise timing relationships between mailbox message transfer and the instruction execution cycles on the two processors. Each invocation of the mailbox would perform only one portion of message testing and transfer, such as test-and-setting a lock variable or transferring one buffer word (or multiple words when simulating direct memory access), tightly interwoven with sub-instruction scheduling of the two processor models. The simple *pollMailbox* procedure shown here and the clock-driven, fine-grain approach discussed represent the boundaries of the range of Tcl-based simulation capabilities possible with Luxdbg.

### 8.4 Demo conclusion

Modular design features of an extension language and periodic refactoring of the design leads to small but powerful extension libraries. We have shown that well engineered extensions allow primitive debug operations at different layers of embedded system abstraction to be combined to support hardware / software co-design and co-debugging.

In particular, this demonstration shows

- a design environment for hardware/software partitioning and rapid prototyping to quickly analyze multiple implementation options
- a debugging environment for software before hardware availability, and
- a test-bench for measuring and verifying system operation.